\documentclass[aps,twocolumn,prl]{revtex4-1}

\usepackage{graphicx}
\usepackage{dcolumn}
\usepackage{bm}
\usepackage{xcolor}
\usepackage{amssymb}   
\usepackage{amsmath}

\newcommand{\muB}{\mu_B}
\newcommand{\la}{\langle}
\newcommand{\ra}{\rangle}
\newcommand{\up}{\uparrow}
\newcommand{\dn}{\downarrow}
\newcommand{\ups}{|\!\!\up\ra}
\newcommand{\dns}{|\!\!\dn\ra}
\newcommand{\beq}{\begin{equation}}
\newcommand{\eeq}{\end{equation}}

\newcommand{\eps}{\varepsilon}
\newcommand{\order}{\mathcal{O}}
\newcommand{\meV}{\text{ meV}}
\newcommand{\pd}{\partial}
\newcommand{\kk}{{\bm k}}
\newcommand{\vg}{{\bm g}}
\newcommand{\re}{{\mathrm{Re}}}   
\newcommand{\im}{\mathrm{Im}}

\begin{document}

\title{Tuning of magnetic quantum criticality in artificial Kondo superlattice CeRhIn$_5$/YbRhIn$_5$}

\author{T. Ishii$^1$}
\author{R. Toda$^1$}
\author{Y. Hanaoka$^1$}
\author{Y. Tokiwa$^{1,2}$}
\author{M. Shimozawa$^3$}
\author{Y. Kasahara$^1$}
\author{R. Endo$^1$}
\author{T. Terashima$^2$}
\author{A. H. Nevidomskyy$^4$}
\author{T. Shibauchi$^5$}
\author{Y. Matsuda$^1$}

\affiliation{$^1$Department of Physics, Kyoto University, Kyoto 606-8502, Japan}
\affiliation{$^2$Research Center for Low Temperature and Materials Science, Kyoto University, Kyoto 606-8501, Japan}
\affiliation{$^3$Institute for Solid State Physics, University of Tokyo, Kashiwa 277-8581, Japan}
\affiliation{$^4$Department of Physics and Astronomy, Rice University, 6100 Main St., Houston, TX 77005, USA}
\affiliation{$^5$Department of Advanced Materials Science, University of Tokyo, Chiba 277-8561, Japan}

\begin{abstract}
The effects of reduced dimensions and the interfaces on antiferromagnetic quantum criticality are studied in epitaxial Kondo superlattices, with alternating $n$ layers of heavy-fermion antiferromagnet CeRhIn$_5$ and 7 layers of normal metal YbRhIn$_5$. As $n$ is reduced, the Kondo coherence temperature is suppressed due to the reduction of effective Kondo screening. The N\'{e}el temperature is gradually suppressed as $n$ decreases and the quasiparticle mass is strongly enhanced, implying dimensional control toward quantum criticality. Magnetotransport measurements reveal that a quantum critical point is reached for $n=3$ superlattice by applying small magnetic fields. Remarkably, the anisotropy of the quantum critical field is opposite to the expectations from the magnetic susceptibility in bulk CeRhIn$_5$, suggesting that the Rashba spin-orbit interaction arising from the inversion symmetry breaking at the interface plays a key role for tuning the quantum criticality in the 
two-dimensional Kondo lattice. 
\end{abstract}
\maketitle

In Kondo lattices consisting of a periodic array of localized spins which are coupled to conduction electrons, a very narrow conduction band is formed at sufficiently low temperatures through the Kondo effect \cite{Hewson}. Such systems are realized in intermetallic heavy-fermion metals, which contain a dense lattice of certain lanthanide (4$f$) and actinide (5$f$) ions. In particular, in Ce(4$f$)-based compounds, strong electron correlations strikingly enhance the quasiparticle (QP) effective mass to about 100 times or more of the bare electron mass, resulting in a heavy Fermi liquid state. In the strongly correlated electron systems, non-Fermi liquid behavior, associated with the quantum fluctuations near a quantum critical point (QCP), a point at which a material undergoes a second-order transition from one phase to another at absolute zero temperature \cite{Sachdev}, has been one of the central issues. The heavy-fermion systems are particularly suitable for this study, because the ground state can be tuned readily by control parameters other than temperature, such as magnetic field, pressure, or chemical substitution \cite{Loh07}. As a result of the many-body effects within the narrow band in these heavy-fermion compounds,  a plethora of fascinating properties have been reported in the vicinity of a QCP.

Recently, a state-of-the-art molecular beam epitaxy (MBE) technique has been developed to fabricate an artificial Kondo superlattice, a superlattice with alternating layers of Ce-based heavy-fermion compounds and nonmagnetic conventional metals with a few atomic layers thick \cite{Shishido}. These artificially engineered materials provide a new platform to study the properties of two-dimensional (2D) Kondo lattices, in contrast to the three-dimensional bulk materials. In the previously studied CeCoIn$_5$/YbCoIn$_5$ superlattices~\cite{Shishido}, where CeCoIn$_5$ is a heavy-fermion superconductor and YbCoIn$_5$ is a conventional metal, each Ce-block layer (BL) is magnetically decoupled from the others, since the Ruderman-Kittel-Kasuya-Yoshida interaction between the spatially separated Ce-BLs is negligibly small due to the presence of the nonmagnetic spacer Yb-BLs. Moreover, the large Fermi velocity mismatch across the interface between heavy-fermion and nonmagnetic metal layers significantly reduces the transmission probability of heavy QPs \cite{She}. In fact, it has been shown that the superconducting heavy QPs as well as the magnetic fluctuations are well confined within the 2D Ce-BLs, as revealed by recent studies of upper critical field and site-selective nuclear magnetic resonance \cite{Mizukami,Yamanaka}. Quantum fluctuations are expected to be more pronounced in reduced spatial dimensions \cite{Monthoux}, and the artificial Kondo superlattices therefore have an advantage to extend the quantum critical regime without long-range ordering. On the other hand, although the previous studies on CeCoIn$_5$/YbCoIn$_5$ superlattices pointed out the importance of the interface between the heavy-fermion and the adjacent normal-metal BLs \cite{Yanase,Goh,Shimozawa}, the question of how the interface affects quantum critical phenomena still remains largely unexplored.  

Here, to study the magnetic quantum criticality of 2D Kondo lattices, we have fabricated superlattices of CeRhIn$_5(n)$/YbRhIn$_5(7)$, formed by alternating layers of heavy-fermion CeRhIn$_5$ \cite{Hegger,TakeuchiMS,Knebel} and normal metal YbRhIn$_5$ \cite{Bukowski}. Bulk CeRhIn$_5$ shows a long-range antiferromagnetic (AFM) order below $T_N=3.8$\,K at ambient pressure; it orders in an incommensurate magnetic structure with ordering vector ${\bm q}=(1/2,1/2,0.297)$ \cite{Bao}. The magnetic order is suppressed by applying pressures  and the ground state becomes purely superconducting at $p>p^\ast\approx1.95$\,GPa with most likely $d$-wave symmetry, similarly to CeCoIn$_5$ \cite{Izawa} and CeIrIn$_5$ \cite{Kasa08}. Compared to the previously studied CeIn$_3$/LaIn$_3$ superlattices \cite{Shishido} based on the cubic heavy-fermion antiferromagnet CeIn$_3$ with higher $T_N=10$\,K and ${\bm q}=(1/2,1/2,1/2)$ \cite{Lawrence}, the effect of reduced dimensionality would be more prominent in the present superlattices based on the tetragonal CeRhIn$_5$ with lower $T_N$. We show that the reduced dimensionality, achieved by reducing $n$, leads to the appearance of the QCP at $n\approx3$. Remarkably, quantum fluctuations in the $n=3$ superlattice are sensitive to the applied magnetic field and its direction. Based on these results, we discuss the significant effect of the inversion symmetry breaking at the interface on the magnetic quantum criticality in 2D Kondo lattice. 

\begin{figure}[t]
	\begin{center}
		\includegraphics[width=0.75\linewidth]{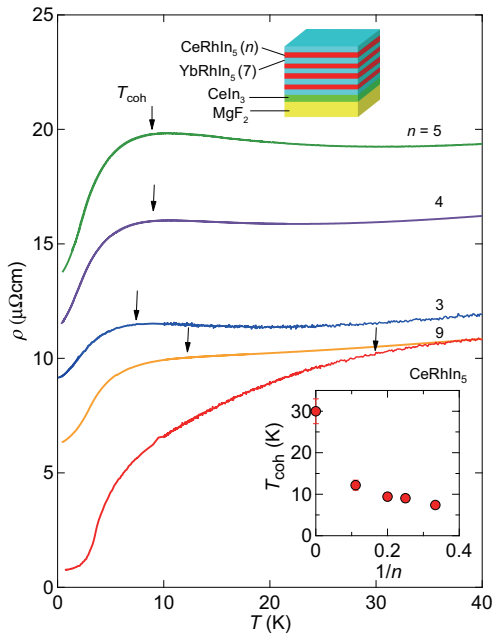}
		\caption{(Color online) 
		Temperature dependence of resistivity $\rho$ for CeRhIn$_5$ thin film and CeRhIn$_5(n)$/YbRhIn$_5(7)$ superlattices. The arrows indicate the Kondo coherence temperature $T_{\rm coh}$. Insets: (top) A schematic representation of the superlattice. (bottom) $T_{\rm coh}$ as a function of $1/n$. 
		}
	\end{center}
\end{figure}

The CeRhIn$_5$($n$)/YbRhIn$_5$(7) Kondo superlattices used for this work were epitaxially grown on MgF$_2$ substrate using the MBE technique \cite{Shishido}. We first grew CeIn$_3$ ($\sim20$\,nm) as a buffer layer, on top of which 7 layers of YbRhIn$_5$ and $n$ layers of CeRhIn$_5$ were stacked alternatively, in such a way that the total thickness was about 300\,nm (see inset in Fig.\,1). Figure\,1 shows the temperature dependence of the resistivity $\rho(T)$ of the $n=9$, 5, 4, and 3 superlattices, along with the CeRhIn$_5$ thin film (300\,nm). The resistivity of the thin film reproduces well that of a single crystal. In the thin film and $n=9$ superlattice, the Kondo coherence temperature  $T_{\rm coh}$ is estimated from the maximum in $\rho(T)$ after subtracting the resistivity of nonmagnetic LaRhIn$_5$ \cite{Shishido02MS} to account for the phonon contribution. The $\rho(T)$ data in the $n=5$, 4, and 3 superlattices shows a maximum at $T_{\rm coh}$ without any background subtraction. As shown in the inset of Fig.\,1, $T_{\rm coh}$ decreases with increasing $1/n$, indicating that reduced dimensionality dramatically suppresses the Kondo coherence \cite{Groten}. It is unlikely that the crystal electric field at Ce-site changes significantly in the superlattices, because it is mainly determined by the neighboring Ce, In, and Rh ions. Therefore, the observed suppression of $T_{\rm coh}$ in the superlattices is a many-body effect likely due to the reduction of the effective number of the conduction electrons which participate in the Kondo screening.  

\begin{figure}[t]
	\begin{center}
		\hspace*{-3.5mm}
		\includegraphics[width=1.05\linewidth]{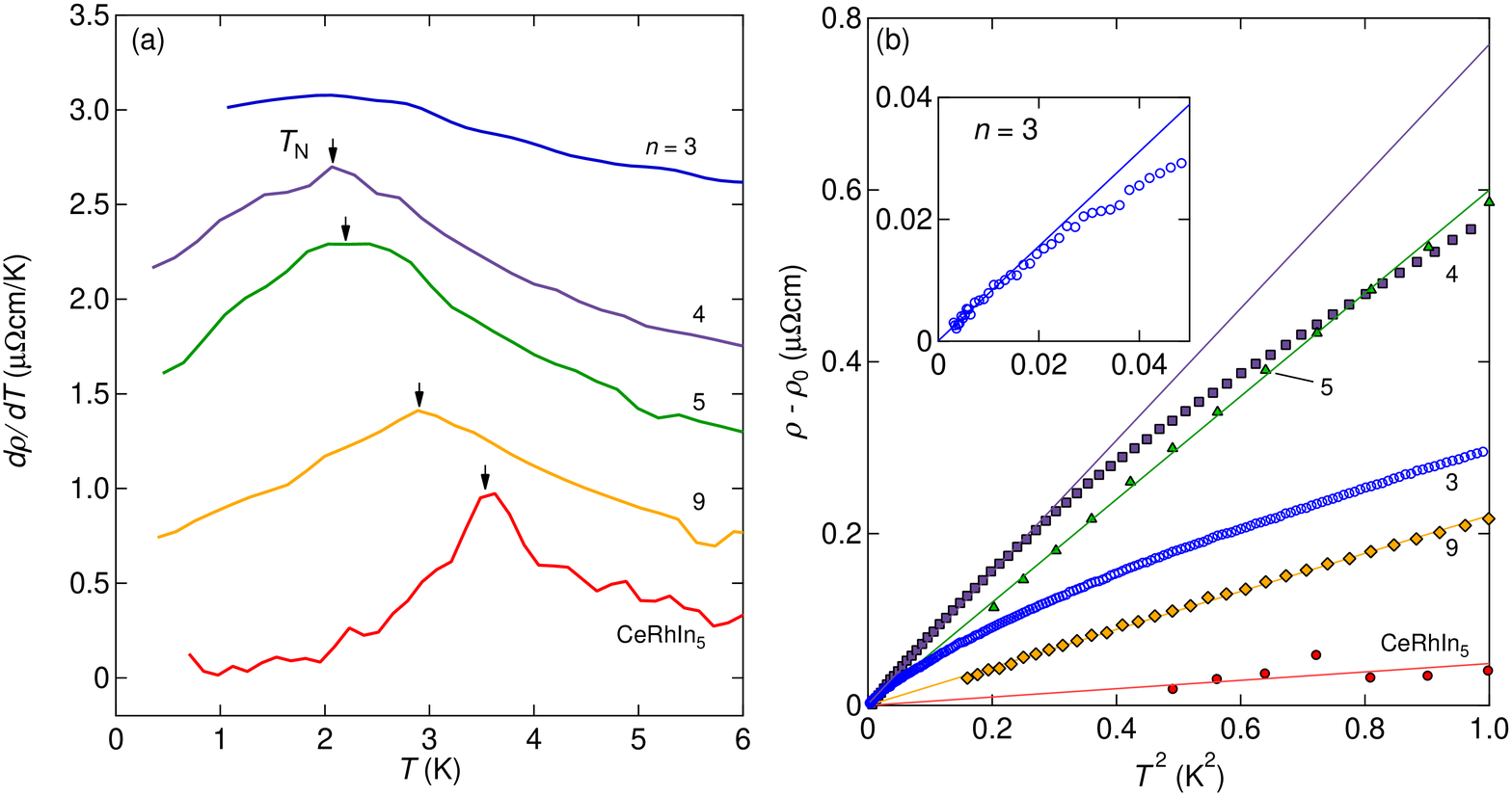}
		\caption{(Color online) 
		(a) Temperature derivative of the resistivity, $d\rho/dT$, as a function of $T$ for CeRhIn$_5$ thin film and CeRhIn$_5(n)$/YbRhIn$_5(7)$ superlattices. Arrows indicate the N\'{e}el temperature $T_N$. 
		(b) $\rho-\rho_0$ plotted against $T^2$. The solid lines are the fits to the $T^2$dependence at the lowest temperatures. The inset displays $\rho-\rho_0$  vs. $T^2$ for the $n=3$ superlattice at low temperatures. 
		}
	\end{center}
	\vspace{-5mm}
\end{figure}

It has been reported that the temperature derivative of the resistivity $d\rho/dT$ exhibits a sharp peak at $T_N$ in the bulk CeRhIn$_5$. Figure 2(a) depicts $d\rho/dT$ at low temperatures for the thin film and the superlattices. In thin film, $d\rho/dT$ reproduces that of the bulk, indicating the AFM transition at $T_N=3.8$\,K \cite{Knebel}. A distinct peak is also observed in the $n=9$, 5, and 4 superlattices, suggesting the presence of the AFM order. For the $n=3$ superlattice, on the other hand, the peak is very broad, and thus, the determination of $T_N$ is ambiguous, which will be discussed later. Obviously, the approach to two dimensions yielded by reducing $n$ enhances quantum fluctuations and thus reduces $T_N$. To see whether the resistivity obeys the Fermi-liquid expression, $\rho=\rho_0+AT^2$, where $\rho_0$ is the residual resistivity and $A$ is the Fermi liquid coefficient, the resistivity at low temperatures is plotted as a function of $T^2$ in Fig.\,2(b). The resistivity of the thin film and $n=9$, 5, and 4 superlattices are well fitted by $\rho\propto T^2$ in a wide temperature range. On the other hand, as shown in the inset of Fig.\,2(b), the $T^2$-dependence is observed only at very low temperatures for the $n=3$ superlattice, consistent with approaching a QCP.    

\begin{figure}[t]
	\begin{center}
		\includegraphics[width=0.95\linewidth]{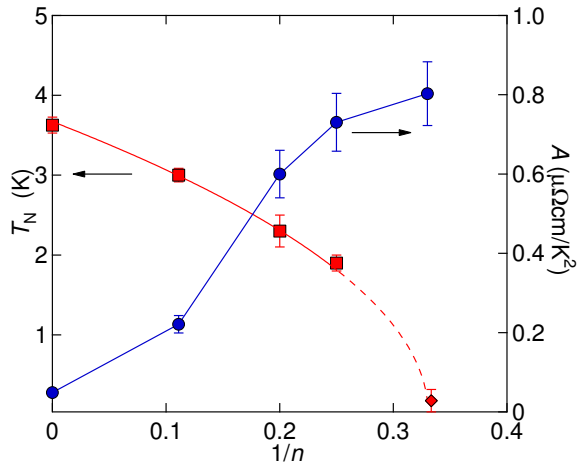}
		\caption{(Color online) 
		The N\'{e}el temperature $T_N$ (left axis) and Fermi liquid coefficient $A$ derived from the expression $\rho=\rho_0+AT^2$ as a function of $1/n$. $T_N$ for the $n=3$ superlattice (diamond) is estimated by the temperature below which the Fermi liquid behavior is observed. 
		}
	\end{center}
\end{figure}

Figure\,3 depicts the thickness dependence of $T_N$ and $A$ that is related to the QP effective mass $m^\ast$ through the effective specific heat coefficient $\gamma_\mathrm{eff}\propto m^\ast$ and the Kadowaki--Woods Fermi liquid relation, $A/\gamma_\mathrm{eff}^2= 1\times 10^{-5}$\, $\mu \Omega$cm(mol K$^2$/mJ)$^2$ \cite{Kontani}. Concomitantly with the disappearance of $T_N$ near $n=3$, $A$ is strikingly enhanced, about 20-fold of its magnitude in the bulk. It is natural to consider that the mass enhancement and the deviation from the Fermi liquid behavior of the resistivity for the $n=3$ superlattice are caused by the quantum critical fluctuations associated with the QCP in the vicinity of $n=3$, i.e. dimensional tuning of the quantum criticality.

To further elucidate the nature of the QCP, we study the magnetoresistance and its anisotropy in an applied magnetic field. Figures\,4(a) and 4(b) show the evolution of $\alpha$, the exponent in $\rho(T)-\rho_0=\Delta\rho(T) \propto T^{\alpha}$, within the field-temperature ($B-T$) phase diagram of the $n=3$ superlattice, for the magnetic field applied parallel to the $ab$ plane and $c$ axis, respectively. For both field directions, the exponent $\alpha$ at low temperatures is strongly affected by the magnetic field \cite{Custers,Gegen,Kasa10}. At $B_c\approx1.2$\,T for ${\bm B}\parallel ab$ and $B_c\approx2$\,T for ${\bm B}\parallel c$, the non-Fermi liquid behavior ($\alpha\lesssim1.5$) is observed down to the lowest temperatures and in a largely extended field range at higher temperatures. For $B>B_c$, a broad crossover regime from the non-Fermi liquid state to the field-induced Fermi liquid state at lower temperature is found to occur. Thus, the non-Fermi liquid behavior dominates over a funnel shaped region of the $B-T$ phase for both field directions. We note that similar phase diagrams have been reported in the bulk heavy fermion compound YbRh$_2$Si$_2$ \cite{Custers}, which constitutes one of the best studied examples of quantum criticality. In Figs.\,4(a) and 4(b), the field dependence of $\gamma_\mathrm{eff}$ for the $n=3$ superlattice estimated from the $T^2$-dependent resistivity in the Fermi liquid regime is also plotted. As the field approaches $B_c$ from either side, $\gamma_\mathrm{eff}$ is rapidly enhanced. These results corroborate the emergence of a field-induced QCP at $B_c$. Although $d\rho/dT$ does not show a discernible peak for the $n=3$ superlattice in Fig.\,2(a), the N\'{e}el temperature is roughly estimated to be $T_N\sim 0.16$\,K by the temperature below which the Fermi liquid behavior is observed (see inset of Fig.\,2(b)). We conclude that in the present 2D Kondo lattice, quantum fluctuations are sensitive to the applied magnetic field; fields of about 1\,T are sufficient to induce a QCP, above which Fermi liquid state with a strongly field-dependent QP mass appears. This small $B_c$ is in sharp contrast to the magnetic field of $\approx50$\,T required to suppress $T_N$ in the bulk CeRhIn$_5$ \cite{Jiao}.

\begin{figure}[t]
	\begin{center}
		\hspace*{-3.5mm}
		\includegraphics[width=1.1\linewidth]{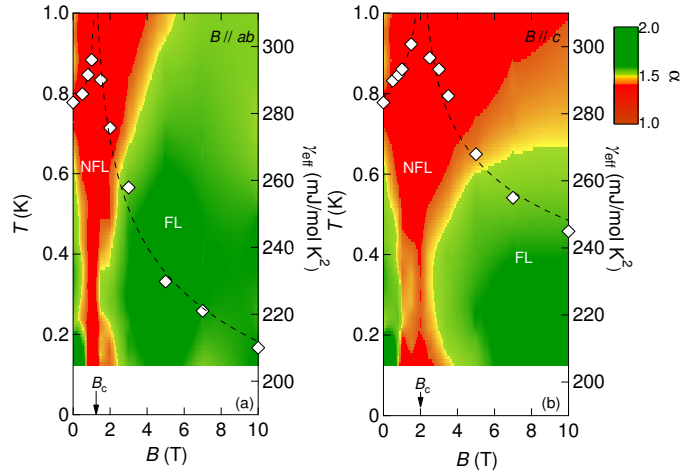}
		\caption{(Color online) 
		Temperature and magnetic field evolution of the exponent $\alpha$ derived from the expression $\rho(T)=\rho_0+AT^\alpha$ in the $n=3$ superlattice (a) for ${\bm B}\parallel ab$ and (b) for ${\bm B}\parallel c$. 
		The triangles represent the effective specific heat $\gamma_\mathrm{eff}$ estimated from the resistivity at the lowest temperatures assuming the $T^2$-dependent resistivity (right axis). 
		}
	\end{center}
	\vspace{-5mm}
\end{figure}

The quantum fluctuations are also sensitive to the field direction. It should be noted that in the bulk CeRhIn$_5$, magnetic susceptibility perpendicular to the $ab$ plane is much larger than that parallel to the plane, $\chi_c\gg\chi_{ab}$ \cite{TakeuchiMS}, implying that the field response for  ${\bm B}\parallel ab$ is expected to be much less sensitive than that for ${\bm B}\parallel c$,  similar to the case of YbRh$_2$Si$_2$ where the magnitudes of $B_c$ for the different field directions are proportional to the magnetic anisotropy.  What is remarkable is that, as seen from Figs.\,4(a) and 4(b), the quantum  fluctuations are more easily suppressed for ${\bm B}\parallel ab$ than for ${\bm B}\parallel c$. Thus, the anisotropy of the quantum critical field in the present 2D Kondo lattice is opposite to that of the bulk susceptibility. Interestingly, recent high-field study of the bulk CeRhIn$_5$ finds that the critical value of $B_c$ at $T=0$ is isotropic~\cite{Jiao}, although it has been reported that the suppression of $T_N$ by magnetic field occurs more rapidly for ${\bm B}\parallel c$ than for ${\bm B}\parallel ab$ at low fields, opposite to what we observe in Fig.\,4.

To explain the observed anisotropy of the critical field, we point out the importance of the space inversion symmetry. Here, in the $n=3$ superlattice, the inversion symmetry is locally broken at the top and bottom layers of the CeRhIn$_5$ blocks in the immediate proximity to the YbRhIn$_5$ layers, whereas the symmetry is preserved in the middle layer \cite{Yamanaka,Yanase,Goh,Shimozawa}. In the absence of inversion symmetry, asymmetry of the potential in the direction perpendicular to the 2D plane $\nabla V \parallel\mathrm{[001]}$ induces Rashba spin-orbit interaction $\alpha_R\,{\bm g}({\bm k})\cdot {\bm \sigma} \propto ({\bm k}\times \nabla V)\cdot {\bm \sigma}$, where ${\bm g}({\bm k})=(k_y,-k_x,0)/k_F$, $k_F$ is the Fermi wave number, and $ {\bm \sigma}$ is the vector of Pauli matrices. The Rashba interaction splits the Fermi surface into two sheets with different spin structures. The energy splitting is given by $\alpha_R$, and the spin direction is tilted into the plane, rotating clockwise in one sheet and anticlockwise in the other \cite{BychkovMS}. Since the noncentrosymmetric interface layers occupy two thirds of the CeRhIn$_5$ layers in the $n=3$ superlattice, the local inversion symmetry breaking at the interfaces, which results in the Rashba spin-orbit splitting of the Fermi surface, has a significant impact on the magnetic properties. In fact, it has been reported that local inversion symmetry breaking strongly affects the superconducting and magnetic properties in CeCoIn$_5$/YbCoIn$_5$ superlattices, leading to the suppression of the Pauli paramagnetic pair breaking effect and magnetic fluctuations at the interface \cite{Yanase,Goh,Shimozawa,Yamanaka}.

In the presence of the local inversion symmetry breaking at the interfaces, the magnetic anisotropy is expected to be modified. For ${\bm B}\perp ab$, the Zeeman splitting $h=g\mu_B J_z B$ enters the energy $\varepsilon_\mathbf{k}$ of QPs at the Fermi level quadratically alongside the Rashba interaction: $E_{\pm}({\bm k}) = \varepsilon_{\bm k}  \pm \sqrt{h^2 + \alpha_R^2 |{\bm g}(\mathbf{k})|^2}$. Therefore, for weak fields ($h\ll \alpha_R$, which is the case here), the Zeeman effect is quadratic rather than linear in field, and is therefore strongly suppressed. By contrast, for in-plane field ${\bm B}\parallel ab$, there is a component of ${\bm B}$ parallel to the Rashba-induced spin ${\bm g}({\bm k})$, and the Zeeman effect is stronger. Therefore, the magnetic susceptibility for ${\bm B}\perp ab$ is expected to be suppressed more strongly than for ${\bm B}\parallel ab$. We theoretically analyzed the anisotropy ratio of the magnetic susceptibility, $\chi_c/\chi_{ab}$, and found that $\chi_c/\chi_{ab}\sim(\chi_c/\chi_{ab})_\mathrm{bulk}\times12(\delta^2+\mathcal{O}(\delta^3))$, where $\delta=h/\alpha_R$. We note that the results are robust against the details of the many-body renormalization of the effective mass (see Supplemental Material \cite{SM}). Realistic values of the Rashba interaction and material parameters of CeRhIn$_5$ lead to the estimate of $\delta$ in the range $(0.02-0.1)$ for field $B=1$ T, yielding the anisotropy ratio $1/100 \lesssim \chi_c/\chi_{ab} \lesssim 1/10$ \cite{SM}, which is opposite to the anisotropy of the bulk CeRhIn$_5$. To confirm this reversed anisotropy of the magnetic susceptibility, more direct measurements which provide microscopic information of the magnetism at the interface, such as site selective nuclear magnetic resonance, are strongly desired.

In summary, to investigate the physical properties of the two-dimensional Kondo lattice, we fabricated CeRhIn$_5(n)$/YbRhIn$_5(7)$ superlattices. As the CeRhIn$_5$ layer thickness is reduced, the effective Kondo screening is largely reduced and the system approaches a quantum critical point in the vicinity of $n=3$. We find that the quantum critical fluctuations, responsible for the non-Fermi liquid behavior, are very sensitive to the applied magnetic field and its direction. The fields of about 1\,T are sufficient to tune the system to the QCP, which is two orders of magnitude smaller than the bulk value. The opposite anisotropy of the quantum critical field between the bulk and $n=3$ superlattice suggests that the Rashba spin-orbit interaction, arising from the local inversion symmetry breaking at the interface, plays an essential role for the magnetism in this artificially engineered 2D Kondo lattice. 

{\it Acknowledgements.} 
We thank Y. Yanase for valuable discussions. 
This work was supported by 
Grants-in-Aid for Scientific Research (KAKENHI) (No.~25220710, 15H02014, 15H02106, 15H05457), and Grants-in-Aid for Scientific Research on innovative areas ``Topological Materials Science" (No.~15H05852) and ``3D Active-Site Science" (No.~26105004) from Japan Society for the Promotion of Science (JSPS). A.H.N. acknowledges the support of the U.S. National Science Foundation grant no. DMR-1350237 and is grateful for the hospitality of the Institute of Solid State Physics, University of Tokyo.

\onecolumngrid
\newpage
\renewcommand{\thefigure}{S\arabic{figure}}
\setcounter{figure}{0}

\section*{Supplemental Material}

\subsection{Fabrication and characterization of the CeRhIn$_5(n)$/YbRhIn$_5(7)$ superlattices}

CeRhIn$_5$/YbRhIn$_5$ superlattices are grown by the molecular beam epitaxy (MBE) technique. The pressure of the MBE chamber was kept at 10$^{-7}$~Pa during the deposition. The (001) surface of MgF$_2$ with rutile structure ($a=0.462$~nm, $c=0.305$~nm) was used as a substrate. The substrate temperature was kept at 550~$^\circ$C during the deposition. Each metal element was evaporated from Knudsen cell for Ce, Yb, and In, and from electron-beam evaporation for Rh. CeIn$_3$ (about 20~nm) was first grown on the MgF$_2$ substrate as a buffer layer. Then, 7-unit-cell-thick (uct) YbRhIn$_5$ and $n$-uct CeRhIn$_5$ layers ware grown alternatively, typically repeated 30-50 times. The deposition rate was monitored by a quartz oscillating monitor and the typical deposition rate was 0.03~nm/s. 

We characterized the quality of the thin films and superlattices by the measurements of the X-ray diffraction (XRD), reflection high-energy electron diffraction (RHEED), atomic force microscopy, and electrical transport measurements. Fig.~S1(a)-(c) shows the XRD patterns for $n=3$, 5, 7, 9 superlattices, 
typical RHEED image, and typical atomic force microscopy image, respectively. 
In the XRD pattern, the positions of the satellite peaks and their asymmetric heights can be reproduced by the step-model simulations ignoring interface and layer-thickness fluctuations, indicating no discernible inter-diffusion across the interfaces. Clear streak pattern in the RHEED image and small surface roughness within 1~nm, comparable to one unit-cell-thickness along the $c$-axis, demonstrate the epitaxial growth of each layer with atomic flatness.

\begin{figure}[h]
	\begin{center}
		\includegraphics[width=0.7\linewidth]{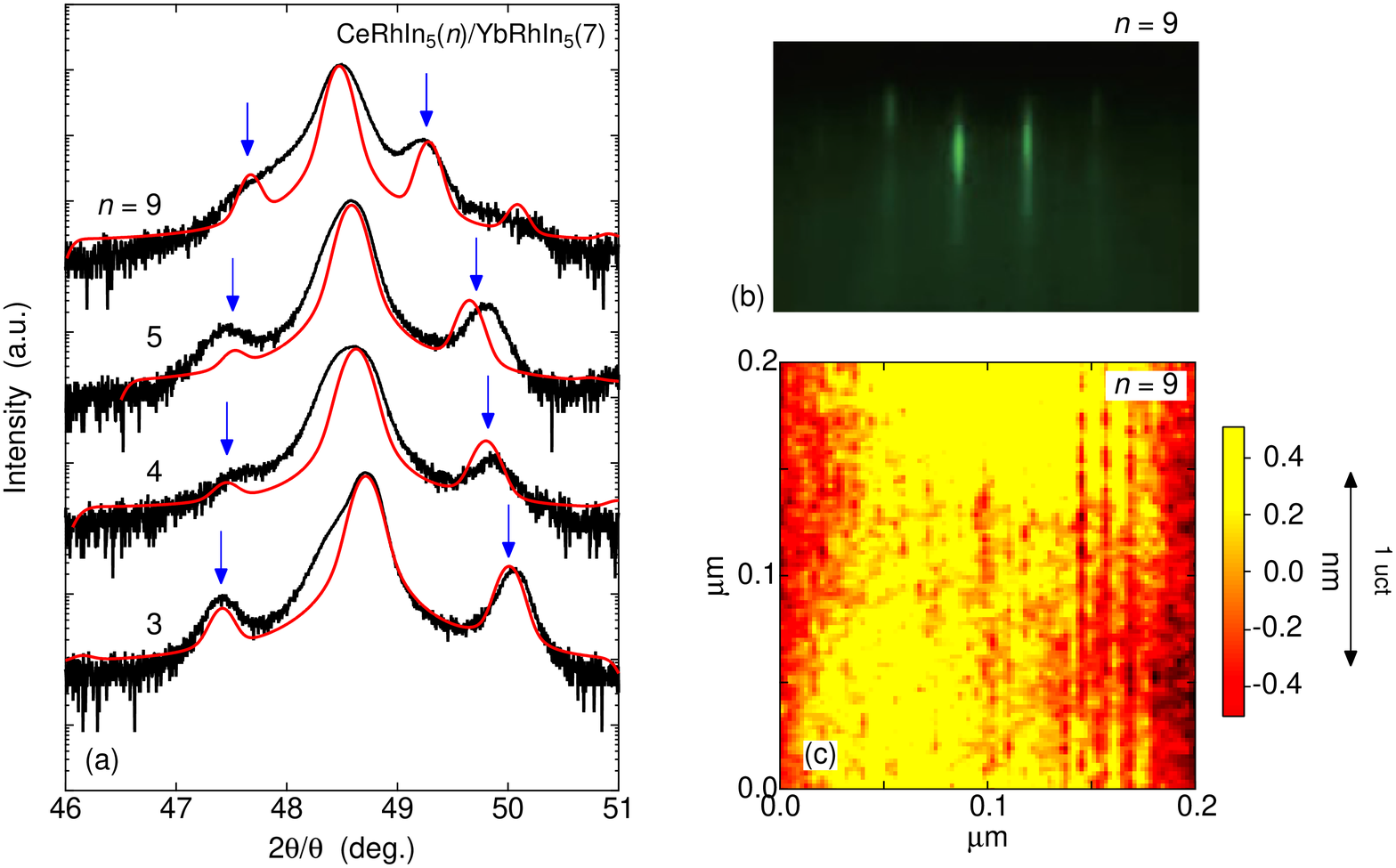}
		\caption{
		(a) Cu K$\alpha_1$ X-ray diffraction patterns for $n = 3$, 4, 5, 9 superlattices around the main (004) peaks. Red lines represent the step-model simulations ignoring the interface and layer-thickness fluctuations. (b) Typical streak patterns of the reflection high-energy electron diffraction image for $n = 9$ superlattice. (c) Typical atomic force microscope image for $n = 9$ superlattice. The surface roughness is within 1 nm, which is comparable to one unit-cell-thickness along the $c$ axis of CeRhIn$_5$.}
	\end{center}
\end{figure}

Fig.~S2(a) shows the temperature dependence of resistivity up to 300~K. The residual resistivity $\rho_0$ as well as $\rho(T=300$~K$)$ shows non-monotonic trend, which presumably comes from uncertainties in determining thickness of the superlattice. We would like to note that, in the superlattice, layer-thickness fluctuation in each block layers are small, because of the presence of clear satellite peak in the XRD pattern (Fig.~S1(a)). Fig~S2(b) shows the residual resistivity ratio, $\rho(T=300$~K$)/\rho_0$, as a function of $1/n$. $\rho(T=300$~K$)/\rho_0$ does not show any clear thickness dependence, implying that no additional disorder is induced by reducing the CeRhIn$_5$ layers.

\begin{figure}[h]
	\begin{center}
		\includegraphics[width=0.8\linewidth]{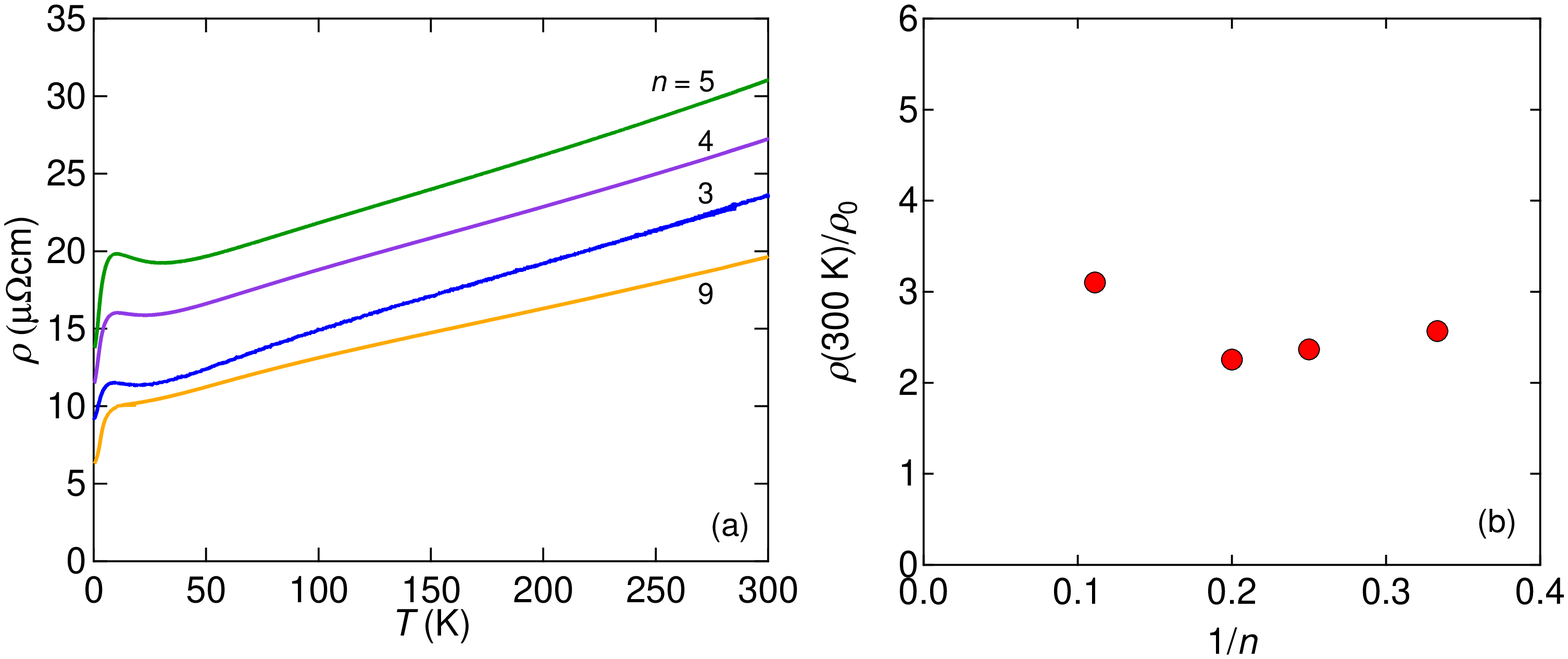}
		\caption{
				(a) Temperature dependence of resistivity $\rho$ for the CeRhIn$_5$($n$)/YbRhIn$_5$(7) 		 superlattices. (b) Residual resistivity ratio, $\rho(300~\mathrm{K})/\rho_0$ as a function of $1/n$.
		}
	\end{center}
\end{figure}

\subsection{Magnetotransport measurements in $n=4$ superlattice}

Fig.~S3(a) and (b) show the temperature ($T$) derivative of the resistivity ($\rho$), $d\rho/dT$, as a function of $T$ for $n = 4$ superlattice in magnetic field parallel to the $ab$ plane ($\mbox{\boldmath{$B$}}\parallel ab$) and $c$ axis ($\mbox{\boldmath{$B$}}\parallel c$), respectively. As explained in the main text of the original manuscript, a peak in $d\rho/dT$ suggests the presence of the antiferromagnetic (AFM) order. The peaks in $d\rho/dT$ are observed in all magnetic fields, and are almost field-independent up to 7 T. This is consistent with the bulk CeRhIn$_5$. Fig.~S3(c) is the same plot as Fig. 4 in the manuscript for $n = 4$ superlattice. At low temperature below 1~K, $\rho(T)$ shows the Fermi liquid behavior with $\alpha = 2$ in $\rho(T)=\rho_0+AT^\alpha$ in the whole field range measured up to 10~T. These results demonstrate that $n = 4$ superlattice show the AFM order and is located far from the quantum critical point (QCP).  

\begin{figure}[h]
	\begin{center}
		\includegraphics[width=0.9\linewidth]{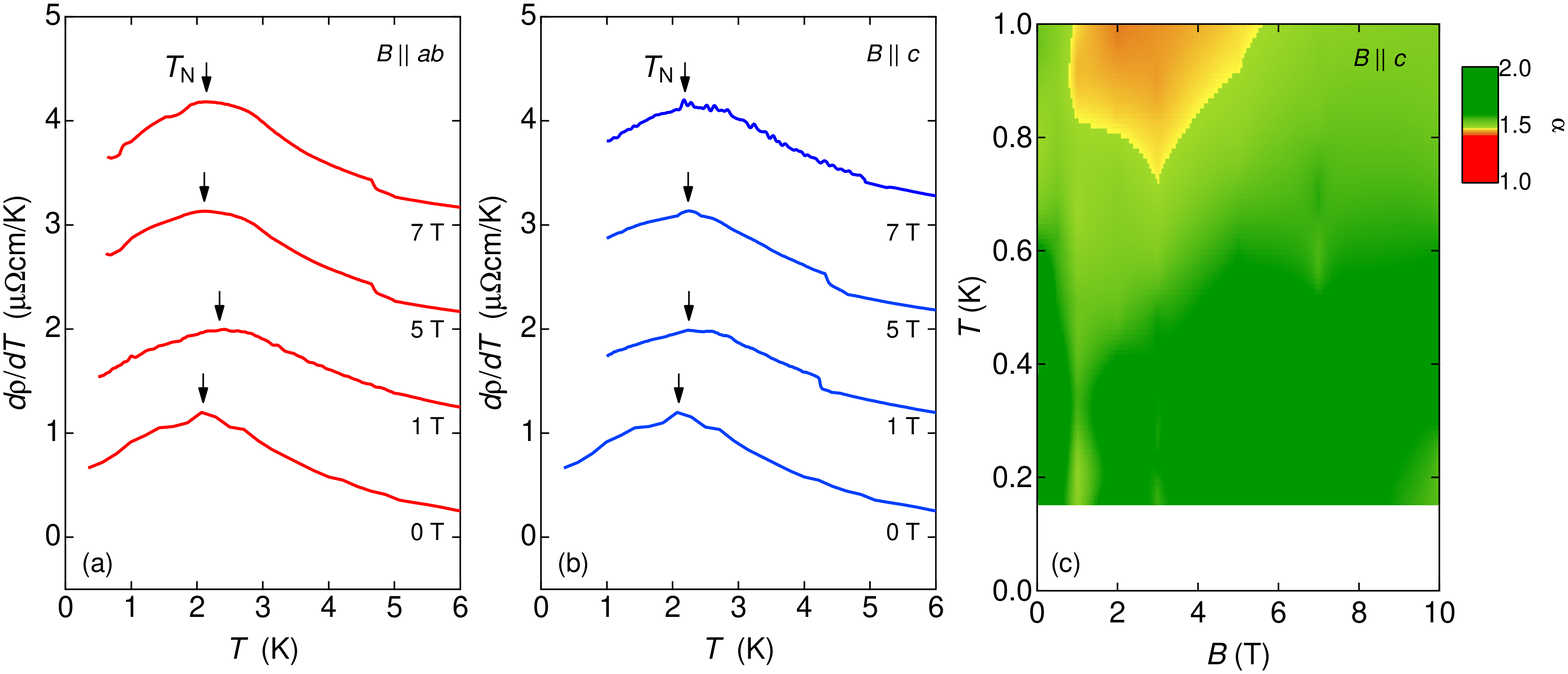}
		\caption{
		Temperature ($T$) derivative of the resistivity, $d\rho/dT$, as a function of T for $n = 4$ superlattices in magnetic field (a) $\mbox{\boldmath{$B$}}\parallel ab$ and (b) $\mbox{\boldmath{$B$}}\parallel c$. (c) Temperature and magnetic field evolution of the exponent α derived from the expression $\rho(T)=\rho_0+AT^\alpha$ for $\mbox{\boldmath{$B$}}\parallel c$ in $n = 4$ superlattice. 
		}
	\end{center}
\end{figure}

\newpage
\subsection{Theoretical analysis of the magnetic field anisotropy}

In bulk CeRhIn$_5$, the anisotropy of the susceptibility is such that $\chi_{zz} > \chi_{xx}$. For instance, in field $B=0.5$~T, the reported magnetization $M_{[001]}=1.5 \muB$ whereas $M_{[100]}=0.80 \muB$ [\onlinecite{Takeuchi2001}], leading to an estimate of the bulk susceptibility anisotropy $\left(\chi_{zz}/\chi_{xx} \right)_{\text{bulk}} \approx 1.9$. However in the CeRhIn$_5(n)$/YbRhIn$_5(7)$ superlattices, our results indicate that the magnetic field response for ${\bm B}\parallel ab$ is stronger than for ${\bm B}\parallel c$ (see Fig. 4 in the main text), in other words the anisotropy appears to be reversed compared to that of the bulk  CeRhIn$_5$. To understand the origin of this phenomenon, we invoke the Rashba spin-orbit coupling associated with the broken inversion symmetry at the interface between CeRhIn$_5$ and YbRhIn$_5$ layers. 

The Rashba interaction originates because of the electric field normal (along [001]) to the heterostructure interface, allowed because of the lack of the inversion symmetry in this direction, and has the form \cite{Rashba}
\beq
H_R = \frac{\alpha}{\hbar}({\bm \sigma} \times {\bm p})_z = \alpha (k_y\sigma_x - k_x\sigma_y) = \alpha_R\, \vg(\kk)\cdot\boldsymbol{\sigma},
\eeq
where $\vg(\kk)=(k_y,-k_x,0)/k_F$ and $\alpha_R = \alpha\, k_F$. The strength of the electric field (absorbed into the constant $\alpha = \alpha_R/k_F$) takes values in the range $(1-10)\times10^{-10}$~eVcm for a number of well studied semiconductor heterostructures (InAs/GaSb \cite{Luo90}, In$_x$Ga$_{1-x}$As/In$_x$Al$_{1-x}$As \cite{Nitta} and GaAs/Al$_x$Ga$_{1-x}$As \cite{Hassenkam97}) depending on the shape of the confining quantum well. For the 2D artificial Kondo superlattices studied here, we provide an estimate of the strength of $\alpha$ towards the end of the Supplemental Material. We note parenthetically that the Dresselhaus spin-orbit effect due to the microscopic crystal field~\cite{Dresselhaus} is typically negligible compared to the Rashba effect, which is why it is sufficient to consider only the latter.

In order to keep the discussion more transparent, we shall first consider the effect of the Rashba interaction on the non-interacting electron gas, before generalizing the results to the more realistic case of heavy Fermi liquid that underpins the electronic properties of CeRhIn$_5$.

\subsection{A. Non-interacting Electron Case}
For transparency, consider first the case of a single non-interacting electron band. Because of the effect of the Rashba interaction, electron spin will precess around the external magnetic field direction ${\bm B}$ as a function of the ${\bm k}$-vector. This can be seen as follows: consider first the field in the $z$-direction perpendicular to the layered superlattice, with the Hamiltonian
\beq
\hat{H} = \eps_{\kk}^{(0)} + h\hat{\sigma}_z + \alpha(k_y\hat{\sigma}_x - k_x\hat{\sigma}_y),
\eeq
where $\eps_{\kk}^{(0)}$ is the bare electron band dispersion and $h = g\muB J_z B$ is the Zeeman field acting on the $z$-component of the angular momentum ${\bm J}$. The corresponding energy eigenvalues $E_{\pm}(\kk) = \eps_{\kk} \pm \sqrt{h^2+\alpha^2 k_\perp^2}$, where $k_\perp = \sqrt{k_x^2 + k_y^2}$ denotes the in-plane momentum. This can be rewritten as
\beq
E_{\pm}(\kk) = \eps_{\kk}^{(0)} \pm \sqrt{h^2+\alpha_R^2 |\vg(\kk)|^2},
\eeq
where for electrons at the Fermi level $|\vg(|\kk|=k_F)|=1$ and hence $\alpha_R$ has the meaning of the Rashba energy splitting in zero external field. The corresponding eigenfunctions yield the following spinor form in the basis of $\ups$ and $\dns$ states (eigenstates of $J_z$): 
\beq
\vert\psi_{+}(k) \ra = \frac{1}{\sqrt{1+b_k^2}}\left(\begin{array} {c}
              1 \\
              b_k 
              \end{array} \right);\qquad 
\vert\psi_{-}(k) \ra = \frac{1}{\sqrt{1+b_k^2}}\left(\begin{array} {c}
              b_k \\
              -1 
              \end{array} \right),
\eeq 
where $b_k = \frac{\alpha(k_y-ik_x)}{h+\sqrt{h^2+\alpha^2 k_\perp^2}}$, manifesting spin precession as a function of $k$. As a result of this spin precession, the $z$-axis magnetization will be $\kk$-dependent: 
\beq
m_z(k) \equiv \frac{1}{2} \left[\la \psi_{+} |J_z| \psi_{+} \ra - \la \psi_{-} |J_z| \psi_{-} \ra\right] = \frac{1-|b_k|^2}{1+|b_k|^2}. 
\label{eq.mz}
\eeq
The zero temperature magnetization along $\hat{n}$
-axis is obtained by summing over the occupied bands:
\beq
M_{n} = (g\muB J_n)\sum_k\sum_\sigma \sigma\, m_n(k)\, \theta(\mu - E_{\kk\sigma}),\quad\quad \left(n=x,\,y,\, z\right).
\label{eq.mag}
\eeq
The expression for the magnetic susceptibility $\chi_n = \pd M_n/\pd B$ is then obtained by differentiating the above equation:
\beq
\chi_n = (g\muB J_n)\frac{\pd M_n}{\pd h_n} = (g\muB J_n)^2 \sum_k\sum_\sigma \sigma\, m_n(k)\,  \left(\frac{-\pd E_{\kk\sigma}}{\pd h_n}\right) \delta(\mu - E_{\kk\sigma}),
\label{eq.chi1}
\eeq
where in the first equality we took into account that Zeeman field $h_n=g\muB J_n B$. Note that in the absence of the Rashba interaction, $m_z$ reduces to 1 in Eq.~(\ref{eq.mz}), and the corresponding magnetization $M_z$ in Eq.~(\ref{eq.mag}) takes on the familiar form $M_z(\alpha=0) = M_{\up} - M_{\dn}$. 

Consider now the field in plane, say, along the $x$-direction. Then the eigenvalues are given by $E_{\pm}(\kk) = \eps_\kk \pm \sqrt{(h+\alpha k_y)^2 + \alpha^2 k_x^2}$. In that case, the spin precesses in $xy$-plane and one can show that the projection of magnetization onto the field direction is given by
\beq
m_x(k) = \frac{h+\alpha k_y}{(h + \alpha k_y)^2 + \alpha^2 k_x^2}.
\eeq   

In order to calculate the susceptibility in Eq.~(\ref{eq.chi1}), we must evaluate the energy derivatives $(\pd E_{\kk\sigma}/\pd h_n)$, yielding:
\beq
\eta_n(k) \equiv \frac{\pd E_{\kk\sigma}}{\pd h_n}
= \left\{ \begin{array}{cc}
\frac{h}{\sqrt{h^2 + \alpha^2 k_\perp^2}} & \text{for } B\, ||\, z \vspace{2mm}  \\ 
\frac{h+\alpha k_y}{\sqrt{(h+\alpha k_y)^2 + \alpha^2 k_x^2}} & \text{for } B\, ||\, x
\end{array} \right.
\label{eq.eta}
\eeq 
It is useful to introduce a variable $\delta = h/\alpha_R \equiv h/(\alpha k_F)$ which denotes the ratio of the Zeeman and Rashba fields at the Fermi level. Then, rewriting the expressions for $m_n(k)$ in terms of $\delta$, the expressions for the susceptibilities become:
\begin{eqnarray}
\chi_{z} & = & (g\muB J_z)^2 \sum_k \sum_\sigma \sigma\, \left\la \frac{\delta^2}{\sqrt{1+\delta^2}}\cdot \frac{\sqrt{1+\delta^2}+\delta}{1 + \delta\sqrt{1+\delta^2}+\delta^2} \right\ra_{F.S.}  \delta(\mu - \eps_{k\sigma}) \nonumber \\
\chi_{x} & = & (g\muB J_x)^2 \sum_k \sum_\sigma \sigma\, \left\la \frac{(\sin\phi+\delta)^2}{(\sin\phi +\delta)^2 + \cos^2\phi} \right\ra_{F.S.}  \delta(\mu - \eps_{k\sigma}),
\label{eq.chi0}
\end{eqnarray} 
where the $\delta$-function pins the momentum integration to the Fermi surface and $\phi=\arctan(k_y/k_x)$ is the azimuthal angle. The sought magnetic field anisotropy is then obtained as a ratio $\chi_z/\chi_x$.

\subsection{B. Susceptibility of a Heavy Fermion System}
The above considerations, obtained for non-interacting electrons, can be generalized to the case of heavy fermion systems. Here, we largely follow the approach developed by Yamada and Yosida [\onlinecite{Yamada1986}] to describe the heavy Fermi liquid. Namely, the Hamiltonian consists of the hybridization $V_{\kk}$ between conduction electrons with dispersion $\eps_{\kk\sigma}$ and $f$-electron band with the dispersion $E_{\kk\sigma}$. This results in a renormalized heavy band with the dispersion $E_{\kk\sigma}^\ast$, obtained in a standard way from the pole of the Green's function:
\beq
G^{-1}(z,\kk) = \left(
\begin{array}{cc} 
z-E_{\kk\sigma}-\Sigma_{\kk\sigma}(z) & -V_\kk \\
-V_\kk^\ast & z-\eps_{\kk\sigma}  
\end{array}
\right),
\eeq 
resulting in the equation for a complex pole $z=E_{\kk\sigma}^\ast - i\,\Gamma_\kk^\ast$ determined from
\beq
(z-E_{\kk\sigma} -\Sigma_{\kk\sigma})(z-\eps_{\kk\sigma})  - |V_\kk|^2 =0,
\eeq
where $\Sigma_{\kk\sigma}$ is the self-energy~\cite{Yamada1986}.

Equation (\ref{eq.chi1}) for the susceptibility remains unchanged, except with the replacement of the non-interacting band $E_{\kk}$ by a heavy electron dispersion $E_{\kk\sigma}^\ast = \re(z)$. Care has to be taken when calculating the derivative $\pd E_{\kk\sigma}^\ast / \pd h_n$ since it involves the derivatives of the self-energy $\pd \Sigma_{\kk\sigma}/\pd h_n$. Here we only quote the final result [see Ref.~\onlinecite{Yamada1986} for details]:
\beq
\frac{\pd E_{\kk\sigma}^\ast}{\pd h_{n\sigma}} = \sigma\,\eta_n(k) \, \left[z_\kk^f\, \tilde{\chi}(\kk) + z_\kk^c\, \frac{g_c \muB/2}{g_f\muB J_n}  \right], \quad (\sigma=\pm 1) \label{eq.pdE}
\eeq
where $g_c$ is the conduction electron $g$-factor with spin 1/2 (different from the $f$-electron $g_f$)  and $h_\sigma = g_f \muB J_n B$ is the Zeeman field felt by the $f$-electrons. Here $\eta_n(k) = \pd E_{\kk\sigma}/\pd h_{n\sigma}$ is the derivative calculated earlier in Eq.~(\ref{eq.eta}) for two different field directions, and $\tilde{\chi}(\kk)$ denotes the non-trivial contribution to the susceptibility coming from the self-energy derivatives:
\beq
\tilde{\chi}(\kk) = 1 - \left.\frac{\pd \Sigma_{\sigma}(\kk,0)}{\pd h_\sigma}\right\vert_{B=0} + \left.\frac{\pd \Sigma_{\sigma}(\kk,0)}{\pd h_{-\sigma}}\right\vert_{B=0}. 
\eeq

The final ingredient necessary for the evaluation of Eq.~(\ref{eq.pdE}) are the quasi-particle residues $z_k^f$ and $z_k^c$, which are defined via the renormalized Green's functions for the localized and conduction electrons, respectively:
\begin{eqnarray}
G^f(\kk,\omega + i\delta) &= & \left[z-E_{\kk\sigma}-\Sigma_{\kk\sigma}(z) - \frac{|V_\kk|^2}{z-\eps_{\kk\sigma}}\right]^{-1} = z_k^f \frac{1}{\omega - E_{\kk}^\ast + i\Gamma_\kk^\ast} \\
G^c(\kk,\omega + i\delta) &= & \left[z-\eps_{\kk\sigma}- \frac{|V_\kk|^2}{z-E_{\kk\sigma}-\Sigma_{\kk\sigma}(z)}\right]^{-1} = z_k^c \frac{1}{\omega - E_{\kk}^\ast + i\Gamma_\kk^\ast}
\end{eqnarray}
with $\Gamma_\kk^\ast = z_k^f\, \im(\Sigma(\kk,\omega+i\delta))$. The density of states at the Fermi level for these electrons is obtained in a standard way as
\begin{eqnarray}
\rho_\kk^f(\omega=0) &=& -\frac{1}{\pi}\im[G^f(\kk,\omega=\mu)] = z_\kk^f \delta(\mu-E_\kk^\ast) \\
\rho_\kk^c(\omega=0) &=& -\frac{1}{\pi}\im[G^f(\kk,\omega=\mu)] = z_\kk^c \delta(\mu-E_\kk^\ast) 
\end{eqnarray}

Putting all the ingredients together, the susceptibilities become
\begin{eqnarray}
\chi_{z} & = & (g\muB J_z)^2 \sum_k \sum_\sigma \sigma\, \left\la \frac{\delta^2}{\sqrt{1+\delta^2}}\cdot \frac{\sqrt{1+\delta^2}+\delta}{1 + \delta\sqrt{1+\delta^2}+\delta^2} \right\ra_{F.S.}   \left[\rho_\kk^f(0)\,\tilde{\chi}(\kk) + \rho_\kk^c(0)\,\frac{g_c \muB/2}{g_f\muB J_z}\right] \nonumber
 \\
\chi_{x} & = & (g\muB J_x)^2 \sum_k \sum_\sigma \sigma\, 
\left\la \frac{(\sin\phi+\delta)^2}{(\sin\phi +\delta)^2 + \cos^2\phi} \right\ra_{F.S.}  
\left[\rho_\kk^f(0)\,\tilde{\chi}(\kk) + \rho_\kk^c(0)\,\frac{g_c \muB/2}{g_f\muB J_z}\right]. 
\label{eq.chi}
\end{eqnarray} 
Comparing these expressions with the non-interacting analogs in Eqs.~(\ref{eq.chi0}), we note that while the interactions renormalize in a very non-trivial way the electronic density of states, the key observation is that the Zeeman--Rashba ratio $\delta = h/\alpha_R$ enters these expressions in the same way as for the non-interacting case. Therefore, while the accurate calculation of the susceptibilities is very complicated and depends on the details of the hybridization and many-body self-energy, the ratio $(\chi_z/\chi_x)$, in the first approximation, only depends on $\delta$. Strictly speaking, this statement is only valid provided the quasi-particle residues ($z_{\kk}^f$  and $z_{\kk}^c$) are isotropic over the Fermi surface, which may not be true in a realistic system. Nevertheless, this is a reasonable approximation under which one can easily evaluate the ratio of the susceptibilities as shown below.

\subsection{C. {$\text{CeRhIn}_5(n)/\text{YbRhIn}_5(7)$} Heterostructures: Anisotropy}
The Fermi surface averages in Eq.~(\ref{eq.chi}) are straightforward to calculate in the limit of the cylindrical Fermi surface in the $xy$-plane, which is justified by the quasi-two-dimensional band structures of the heterostructures and agrees with the quantum oscillation data for CeRhIn$_5$~\cite{Shishido2002}. This yields the anisotropy ratio: 
\beq
\frac{\chi_z}{\chi_x} \sim \left(\frac{J_z}{J_x}\right)^2\; \frac{4\delta\,(\delta + 2\delta^2 + \order(\delta^3))}{1/3 + \frac{2}{15}\delta^2 + \order(\delta^4)} \approx \left(\frac{\chi_z}{\chi_x} \right)_{\text{bulk}} \times 12 (\delta^2 + 2\delta^3 + \order(\delta^4)),\quad\quad \delta\ll 1
\label{eq.ratio}
\eeq
The above result is valid in the regime $\delta \ll 1$ (i.e. $h\ll \alpha_R$), which we show below holds in these CeRhIn$_5$ heterostructures. In the opposite limit of vanishing Rashba coupling ($\alpha=0$), the Fermi surface averages yield 1, as expected, and we thus obtain the bulk ratio $\left(\chi_z/\chi_x \right)_{\text{bulk}} = (J_z/J_x)^2 \approx 1.9$.

Let us now proceed to estimate the value of $\delta$, which can be written as follows:
\beq
\delta = \frac{h}{\alpha k_{F\perp}} = \frac{1}{2} \frac{h}{\alpha k_R/2}\,\frac{k_R}{k_{F\perp}} = \frac{1}{2}\frac{h}{E_R}\,\frac{k_R}{k_{F\perp}} = \frac{1}{2}\frac{h}{E_R} \sqrt{\frac{E_R}{E_F}},
\eeq
where $E_R = \hbar^2 k_R^2/2m^\ast = \alpha k_R/2$ is the Rashba energy commonly associated with the Rashba vector $k_R = \alpha m^\ast/\hbar^2$. Presumably, the strength of the electric field at the interface is not dissimilar to other semiconductor heterostructures, such as the well studied case of InGaAs/InAlAs. Taking the value of $\alpha=7\times10^{-10}$~eVcm, as determined for the InGaAs/InAlAs interface in Ref.~[\onlinecite{Nitta}], and the effective mass to be that of bulk CeRhIn$_5$ ($m^\ast\approx 6 m_0$ from quantum oscillations~\cite{Shishido2002}), we obtain $E_R \approx 1$~meV. Next, we estimate $E_F\approx \hbar^2k_F^2/2m^\ast$ from the typical de Haas--van Alphen frequency $F\sim1$~kT \cite{Shishido2002}, which using Onsager's relation, yields $\pi k_F^2 = (2\pi e/\hbar)F$. This translates into the Fermi energy
\beq
E_F = \frac{\hbar^2 k_F^2}{2m^\ast} = 
2\frac{m_e}{m^\ast}\left(\frac{e\hbar}{2m_e}\right)F = 2 \left(\frac{m_e}{m^\ast}\right) (\muB F)
\eeq
about 
$E_F \approx 20$~meV. 
Therefore, taking the Zeeman energy in the field of $\mu_0 H = 1$~T to be $h \sim 1$~K, we estimate
\beq
\delta = \frac{1}{2}\frac{h}{E_R}\,\sqrt{\frac{E_R}{E_F}} \sim 
\frac{0.1\meV}{1\meV}\sqrt{\frac{1 \meV}{20 \meV}} \approx \frac{1}{50}.
\eeq
Substituting into Eq.~(\ref{eq.ratio}), we find the anisotropy ratio of susceptibilities
\beq
\frac{\chi_z}{\chi_x} \approx \frac{1}{100}, 
\eeq
and we conclude that the presence of Rashba interaction strongly renormalizes the anisotropy by suppressing the $z$-axis susceptibility. 

The calculated ratio likely overestimates the Rashba coupling strength. However even if we take $E_R$ to be 10 times smaller (of the order 0.1~meV), we still find the large anisotropy $\chi_z/\chi_x \approx 1/10$ due to the smallness of the ratio $E_R/E_F$. We conclude that these estimates are therefore robust and the anisotropy ratio should become even more pronounced in low fields, with the electron spins essentially confined to the easy $xy$-plane because of the Rashba spin-orbit coupling.

Note that, as follows from the above estimates, the Zeeman splitting and Rashba energy are negligible compared to the electron bandwidth, thus justifying \emph{a posteriori} us taking the densities of states to be at the Fermi level in Eqs.~(\ref{eq.chi}).

\end{document}